# EDU-Net: Retinal Pathological Fluid Segmentation in OCT Images with Multiscale Feature Fusion and Boundary Optimization

Zijun Lei[1,2,3,a], Zikang Xu[4,5,a], Liang Zhang[4,5], Ge Song[1,2,3], Hanyu Guo[1,2,3]
Dan Cao[4,5,*], Yujia Zhou[1,2,3,*], Qianjin Feng[1,2,3,*]

[1] School of Biomedical Engineering, Southern Medical University, Guangzhou, 510515, China
[2] Guangdong Provincial Key Laboratory of Medical Image Processing, Southern Medical University, Guangzhou, 510515, China
[3] Guangdong Province Engineering Laboratory for Medical Imaging and Diagnostic Technology, Southern Medical University, Guangzhou, 510515, China
[4] Guangdong Cardiovascular Institute, Guangdong Provincial People's Hospital, Guangdong Academy of Medical Sciences, Guangzhou 510000, China.
[5] Department of Ophthalmology, Guangdong Eye Institute, Guangdong Provincial People's Hospital (Guangdong Academy of Medical Sciences), Southern Medical University, Guangzhou 510000, China.

* Corresponding author
caodan@gdph.org.cn (D. Cao); yujia90@smu.edu.cn (Y. Zhou); fengqj99@smu.edu.cn (Q. Feng)
[a] These authors contributed equally to this work.

**Financial Support:** This work was supported in part by the National Natural Science Foundation of China [grant number 82371063], the National Natural Science Foundation of China (No. 62471214, 62276122, and 12126603), the Natural Science Foundation of Guangdong Province, China (No.2023A1515011956), the Natural Science Foundation of Guangdong Province, China (No.2024A1515012291), and the Guangzhou Science and Technology Project, China (No. 2024A04J9911).

**Conflict of Interest:** The authors declare that they have no known competing financial interests or personal relationships that could have appeared to influence the work reported in this paper.

**Running Head:** EDU-Net: Multi-scale and Boundary Optimization for OCT Segmentation

# ABSTRACT

**Objective:** Diabetic macular edema (DME) is the leading cause of severe visual impairment in patients with diabetes. Quantification of retinal fluid, particularly intraretinal fluid (IRF) and subretinal fluid (SRF), plays a critical role in the management of DME. Although optical coherence tomography (OCT) can be used for detection, the variable morphology of fluid accumulation and the blurred boundaries caused by noise interference still limit the accuracy of OCT's automatic segmentation.

**Design:** Retrospective model development and validation study.

**Subjects:** We first pre-trained the model on 700 OCT images selected from the Mendeley data, then fine-tuned and tested it using the in-house dataset, and conducted five-fold cross-validation on the RETOUCH dataset.

**Methods:** This study proposes a novel edge-guided dual-branch encoder-decoder network (EDU-Net) to achieve accurate and efficient automatic segmentation of OCT liquid lesions. The local feature extraction branch is based on the EfficientNet model, which precisely captures tiny lesions by leveraging its lightweight separable convolution and high-resolution feature preservation strategy. The global feature extraction branch is based on the large-kernel efficient convolution (LKEC) module and the downsampling layer design to enhance long-range dependencies and global semantics. EDU-Net applies a multi-category edge-guided attention module to fuse high-frequency boundary detail information to each resolution feature to optimize the boundary segmentation performance.

**Main Outcome Measures:** We evaluated model performance primarily by the Dice similarity coefficient (DSC) and demonstrated robustness across datasets.

**Results:** Extensive results on the in-house and public datasets demonstrate that EDU-Net achieves state-of-the-art DSC segmentation performance in terms of efficiency and robustness, especially in the segmentation of IRF lesions.

**Conclusions:** EDU-Net integrates local details with global context and optimizes boundaries, achieving an improvement in the accuracy of automatic segmentation of retinal fluid.

# 1. Introduction

Diabetic retinopathy (DR) is a common microvascular complication closely related to diabetes and is one of the main causes of preventable blindness in the working-age population.[1] The vision impairment caused by DR mainly involves two pathological processes: diabetic macular edema (DME) and proliferative diabetic retinopathy (PDR). DME is a specific type of macular edema caused by damage to the blood–retinal barrier due to diabetes.[2] Timely and precise diagnosis and management of DME are critical for controlling retinal fluid accumulation and inflammatory response. Effective treatment of DME not only improves or stabilizes visual function but may also reduce the likelihood of transition to sight-threatening PDR. Optical coherence tomography (OCT) is widely used for the diagnosis and follow-up of DME.[3] OCT uses near-infrared light to scan the retinal cross-section to display the structure of retinal layers (e.g., nerve fiber layer, outer nuclear layer, and retinal pigment epithelial layer) and abnormal fluid accumulation at high resolution.[4] Accurate quantification of retinal fluid (such as intra fluid [IRF] and subretinal fluid [SRF]) on OCT enables clinicians to better characterize disease severity, tailor individualized treatment regimens, and monitor therapeutic response over time.[5,6] However, manual delineation of complete retinal fluid structure is highly time-consuming and laborious, and inconsistency in the delineation between clinical experts is inevitable. To this end, designing an accurate system that can automatically segment the retinal fluid from background is crucial.

In recent years, deep learning (DL) based on convolutional neural networks (CNNs) has been widely used in various segmentation tasks owing to its state-of-the-art representation performance.[7] However, accurate automatic retinal fluid segmentation remains challenging due to the following factors (Fig. 1):

• **Complex and varied retinal fluid lesion morphology:** Effusions range from micrometer-scale vesicles to millimeter-scale cavities, exhibiting highly heterogeneous size and shape distributions. IRF is more common in DME, and it appears as scattered small cysts or irregular low-reflectivity areas, mostly in the inner nuclear layer. Meanwhile, SRF occurs relatively infrequently and appears as larger areas of continuous fluid.

• **Blurred boundaries:** The speckle noise and low contrast in OCT images yield weak, discontinuous edge gradients around fluid accumulation, leading to ambiguous contours. Therefore, most existing methods may misjudge the boundaries, causing missed detection or over-segmentation.

• **Ineffective fusion of local details and global context:** Expanding the receptive field via down-sampling may lose local detail information (e.g., U-Net). Relying on shallow networks for local refinement amplifies noise sensitivity and increases false-positive rates (e.g., DeepLabv3+).

## 1.1. Related work

### 1.1.1. Traditional methods for retinal fluid segmentation

Traditional retinal fluid segmentation methods design various rules on the basis of the geometric characteristics of fluid regions. These methods include contour-based[8] and level set.[9] Carass et al.[8] proposed a multi-object geometric deformable model that achieves unsupervised segmentation by simultaneously evolving multiple level-sets of boundaries. Somfai et al.[9] measured the complexity of boundaries on the basis of fractal dimension and combined morphological operations to extract liquid features for automatic segmentation.

Meanwhile, the segmentation problem is usually transformed into a classification/regression problem. On the basis of robust feature extraction methods, such as local binary pattern[10] and histogram of oriented gradients,[11] classifiers like support vector machine[10] and random forest[12] are then applied to classify the features and thus identify lesion areas. Lang et al.[13] achieved automatic segmentation of OCT image through image normalization, retinal boundary detection, multiscale feature extraction and random forest classification. Chiu et al.[14] employed kernel regression denoising and adaptive derivative estimation to extract features, combined weighted sequential forward selection to construct a negative Gaussian classifier, and utilized graph theory segmentation to achieve target boundary detection. In addition to the above supervised classification methods, unsupervised classification methods, such as k-means clustering algorithm, are used to identify lesion areas. Pilch et al.[15] initially divided the OCT image grayscale by using k-means clustering and then identified pathological cavities with a k-nearest neighbor classifier.

However, the accuracy and the generalization ability of the abovementioned methods are limited by hand-crafted features and prior knowledge, especially when facing cases from different datasets or variable morphological structures of the same case.

### 1.1.2. DL methods for retinal fluid segmentation

In recent years, DL has demonstrated excellent performance in medical image segmentation tasks owing to its ability to automatically extract hierarchical semantic features and underlying discriminative information.[16] Multiple DL models, such as FCN,[17] U-Net,[错误!未找到引用源。] Seg-Net,[19] Deeplabv3+,[20] and U-Net++,[21] have been used for retinal lesions and layer segmentation tasks.

Liu et al.[22] proposed a multiscale side output and dual attention mechanism to tackle the challenge posed by large-scale variations in the morphology of fluid lesion regions in OCT and presented an enhanced nested U-Net architecture (MDAN-UNet) to segment seven retinal layers and three types of fluid in OCT. Feng et al.[23] proposed the CPF-Net model, which explores global and multiscale contextual information mainly through a soft spatial attention mechanism that learned features across different receptive fields by using dilated convolutions with shared weights. Wang et al.[24] developed the MsTGANet model, a multiscale transformer global attention network for drusen segmentation in retinal OCT images. The model combines non-local and multi-semantic information. Rasti et al.[25] recently proposed a novel self-adaptive and multilevel attention convolutional U-shaped network called RetiFluidNet.

Some other works attempted to improve fluid segmentation performance in terms of edge enhancement. Pappu et al.[26] improved segmentation performance by fusing the preprocessed OCT images with edge maps, extracting features through multiscale convolution, and enhancing the perception of blurred boundaries through an edge attention mechanism. Yu et al.[27] combined a contour branch and a diffusion branch to achieve smoother and more accurate segmentation performance. The network is called PadNet, where the contour branch is used to extract edge information from noise and the diffusion branch extends the lesion region boundary through a distance map.

However, existing methods have difficulty in balancing efficiency and accuracy. Complex multiscale feature extraction/fusion strategies and edge enhancement modules with cumbersome preprocessing and multitask parameter tuning processes lead to high training and debugging costs and limited generalization capabilities.

## 1.2. Overview of the propose method

In this study, a novel edge-guided dual-branch encoder-decoder network (EDU-Net) was designed for automatic and accurate retinal fluid segmentation (Fig. 2). EDU-Net was built on a classic U-shaped structure with global-local dual branches, which can efficiently and robustly extract features and fuse complex and diverse retinal fluid lesion morphologies. Specifically, the local feature extraction branch is based on the EfficientNet model, which extracts rich fine-grained information from shallow layers by leveraging its lightweight separable convolution and high-resolution feature preservation strategy. The global feature extraction branch is based on the large-kernel efficient convolution (LKEC) module and the downsampling layer design to obtain robust long-range and global features. A multi-category edge-guided attention (MC-EGA) module was introduced in the global feature extraction branch, which extracts high-frequency edge information on the basis of Laplacian pyramid and then generates and adaptively fuses edge, foreground, and background attention maps for each scale.

## 1.3. Contributions

The main contributions of this study are summarized as follows:

- A high-accuracy, robust, memory-efficient, and fast retinal fluid segmentation algorithm, named EDU-Net, was introduced and validated on expensive experiments with a public dataset and an in-house dataset. Compared with state-of-the-art traditional or deep learning methods, the proposed EDU-Net exhibited superior segmentation and generalization performance.

- A novel global-local dual-branch architecture was introduced, and different upsampling modules were designed to achieve end-to-end segmentation. Global and local features can be collaboratively modeled to handle complex and varied retinal fluid morphology regions by fusing different branches.

- An MC-EGA module was adopted, and an effective fusion of high-frequency boundary information and global/local semantic features was achieved. The edge response was amplified by adaptively fusing the above information layer by layer to reduce the degradation of segmentation performance caused by blurred boundaries.

## 2. Methods

As illustrated in Fig. 2, EDU-Net was proposed for automatic retinal fluid segmentation, integrating three core components: (1) an efficient local detail information extraction branch based on the classical EfficientNet, (2) a global semantic feature extraction branch by an LKEC module and downsampling layers, and (3) an edge-enhanced strategy using an MC-EGA module.

### 2.1. Local feature extraction branch

The local feature extraction branch is based on the classical EfficientNet to efficiently provide rich detail information by hierarchical structures (Fig. 3). The compound scaling strategy proposed in[28] was adopted to balance the depth, width, and resolution of the network. The local encoder consists of eight stages. The first stage is a $3\times3$ convolution with batch normalization and swish activation, which is used for initial feature extraction and down-sampling. The next seven stages gradually refine the features by stacking a mobile inverted bottleneck convolution (MBConv) module with an inverted residual connection (IRC).

The MBConv module is the core building block of this branch, as shown in Fig. 4(a), and its structure is as follows: first, a $1\times1$ convolution is used to expand the channels, and then a $k\times k$ (typically $3\times3$ or $5\times5$) depth-wise convolution is adopted to extract features, followed by an SE module composed of global average pooling and two fully connected layers to calculate channel attention. Finally, a $1\times1$ convolution is used to reduce the dimension.

In the decoder of the local feature extraction branch Fig. 4(b), upsampling is directly achieved by transposed convolution with a factor of two, followed by BatchNorm + ReLU. After the feature maps of a specific scale obtained from the encoder are concatenated, two $3\times3$ convolutions + BatchNorm + ReLU (double-conv) are used to deeply fuse multiscale features.

## 2.2. Global feature extraction branch

The global feature extraction branch is designed to efficiently achieve global information by large-kernel efficient convolution and downsampling structures. As shown in Fig. 2, a 4×4 non-overlapping convolution with a stride of 2 is first applied at the input end to downsample the original feature map. This step not only significantly expands the initial receptive field but also reduces the subsequent computational cost. Second, the network is composed of multiple LKEC modules connected in series with a 2×2 non-overlapping convolution with a stride of 2 as downsampling

As shown in Fig. 5(a), in each LKEC module, a 7×7 depth-wise separable convolution is first executed to capture large receptive field features, followed by LayerNorm. Second, the number of channels is expanded by 4 times through a 1×1 convolution, activated by GELU, and projected back to the original channel dimension through a 1×1 convolution. The output is then multiplied by a learnable layer scale scaling coefficient and finally added to the input residual through DropPath.

In the decoder of the global feature extraction branch [Fig. 5(b) and (c)], two convolutions are applied to the decoder features to map the number of channels to the number of target categories, generating the "coarse classification" feature map required by the MC-EGA module [Fig. 5(b)] In the upsampling path [Fig. 5(c)], the original decoder features are concatenated with the edge-enhanced feature output by the corresponding MC-EGA module, and the feature map is upsampled by a factor of 2 through bilinear interpolation, followed by a 3×3 convolution + BatchNorm + GELU activation to enhance nonlinear expression capability.

## 2.3. Multi-category edge-guided attention module

The MC-EGA module is designed to effectively preserve multiscale boundary information, thereby reducing the degradation of segmentation performance caused by boundaries blur and strengthening the foreground region by using the multi-category attention mechanism. Notably, the MC-EGA module is equipped on the global feature extraction branch only and not on the local feature extraction branch. On the one hand, ensuring the efficiency of local feature extraction is crucial, adding additional MC-EGA modules could result in high computational cost. On the other hand, high-frequency boundary features refinement of local features could amplify noise sensitivity and increase the false-positive rate.

As shown in Fig. 6, at the $i^{th}$ layer, the MC-EGA module receives three inputs: embedded features from the encoder $f_i^e \in R^{H_i \times W_i \times N_i}$, high-frequency features extracted through the Laplacian pyramid method $f_i^l \in R^{H_i \times W_i \times 1}$, and the predicted features from the $(i+1)^{th}$ layer of the decoder $f_{i+1}^d \in R^{H_i \times W_i \times 1}$. The extraction of high-frequency feature $f_i^l$ employs the Laplacian pyramid method, which uses the difference of Gaussian (DoG) operator to approximate the Laplacian operator for preserving multiscale edge details as follows:

$$f_0^l = I - G(I), \; f_i^l = \text{Downsample}(f_{i-1}^l) \qquad (1)$$

Where $I$ is the input image, and $G(\cdot)$ is the Gaussian blur operation. For the case $i \geq 1$, high-frequency features are obtained through downsampling.

Bilinear interpolation was applied to upsample $f_i^l$ to ensure that the high-frequency features of different scales are consistent in spatial dimensions with the encoder features, thereby obtaining the edge attention as follows:

$$A_{\text{edge}} = \text{Interpolate}(f_i^l, \text{size} = (H, W)) \qquad (2)$$

An activation function was applied to the predicted feature $f_{i+1}^d$ to obtain the probability distribution $p$ for each category to adapt to multi-category segmentation scenarios. Here, the $0^{th}$ category typically represents the background. Thus, the background attention $A_{bg}$ and the foreground attention $A_{fg}$ can be defined as follows:

$$A_{bg} = p[:,0:1,:,:] \tag{3}$$

$$A_{fg} = \frac{1}{C-1}\sum_{c=1}^{C-1} p[:,c,:,:] (\text{or } 1 - A_{bg}) \tag{4}$$

Based on the above attention maps, the encoder feature $f_i^e$ was element-wise multiplied with the background, foreground, and edge attentions. Then, they were concatenated with the original encoder feature $f_i^e$ along the channel dimension. After convolution operations, the combined features can be generated as follows:

$$f_i^a = f_i^e + G([f_i^e \otimes A_{bg}, f_i^e \otimes A_{fg}, f_i^e \otimes A_{edge}]) \tag{5}$$

The function of $G$ is to generate spatial weight masks via convolution and activation functions and then perform weighted fusion on the input features to strengthen the representation of lesion edges. Finally, the convolutional block attention module (CBAM) was employed as follows to recalibrate the attention feature $f_i^a$ to capture subtle feature correlations between the boundary regions and the background regions:

$$f_i^d = \text{CBAM}(f_i^a) \tag{6}$$

## 2.4. Loss function

The following loss function of EDU-Net consists of two parts: the dice similarity coefficient (DSC) loss of the global feature encoding–decoding branch and the local feature encoding–decoding branch:

$$L = \alpha L_{DSC-global} + \beta L_{DSC-local} \tag{7}$$

## 2.5 Dataset

The experiment involved an in-house dataset (collected from Guangdong Provincial People'

s Hospital) and two publicly available datasets (Mendeley data[29] and RETOUCH Challenge[30]). Two training strategies were designed to comprehensively evaluate the model performance. First, pre-training was conducted on a portion of the Mendeley data to extract general features, followed by fine-tuning and testing on the basis of the in-house clinical data to verify the model's applicability in clinical scenarios. Second, further validation was implemented on the RETOUCH Challenge dataset to examine the model's stability and generalization ability. Five-fold cross-validation was adopted to divide the datasets and obtain a reliable performance. Each scan was center cropped and resized to 512×512 due to the different sources of dataset images. The specific information of these datasets is as follows:

### 2.5.1. Mendeley data

Based on a large OCT dataset (11,349 DME images collected by Spectralis scanner) on the Mendeley data platform, 106 confirmed DME patients were selected, totaling 700 images as pre-training data. The original image resolutions covered five types: 512×496, 512×512, 768×496, 1024×496, and 1536×496. All lesion annotations were independently completed by two senior ophthalmologists in Mimics (version 26.0) and reviewed by a third expert to ensure annotation consistency and quality, as shown in Fig. 7. Finally, the images were divided into training and validation sets in a 7:3 ratio for the subsequent model pre-training.

### 2.5.2 Clinical dataset

Clinical data were collected from a local hospital in China (Guangdong Provincial People's Hospital Affiliated to Southern Medical University), including a total of 200 OCT retinal B-scan images covering various disease stages. Ethical approval was granted by the Institutional Review Board of Guangdong Provincial People's Hospital (Approval No. KY2024-065-01), and a written informed consent was obtained from all enrolled participants. All images were acquired using the Spectralis scanner with a horizontal pixel spacing of 5.60 $\mu m$ and a vertical pixel spacing of 3.87 $\mu m$, and the original resolution was 512×496. The annotation method was consistent with that of the pre-training dataset. The original dataset was divided into two subsets in the same proportion in accordance to lesion types. The two subsets were used for model fine-tuning and model testing to comprehensively evaluate the model's performance under clinical conditions.

### 2.5.3 RETOUCH Challenge

The RETOUCH Challenge dataset contains images from three OCT scanners (Zeiss Cirrus, Heidelberg Spectralis, and Topcon), and it has been widely used in related research. As shown in Fig. 7, the image quality of the Cirrus type is relatively low, and it contains an additional lesion label of pigment epithelial detachment (PED), which leads to significant differences from clinical data.

## 3. Experiment configurations

## 3.1 Implementation details

All research processes were carried out on a workstation equipped with an NVIDIA GeForce RTX 2080 Ti graphics card, based on a Python 3.8 and PyTorch 2.4.1+cu121 environment. The model training was set with a maximum epoch of 100, an initial learning rate of 1e-4, and a batch size of 4. The Adam optimizer was used, and the learning rate was dynamically and adaptively adjusted on the basis of the validation set loss through the ReduceLROnPlateau scheduler. Various data augmentation techniques were applied in the training mode, including random horizontal flipping, rotation, and adjustments to brightness and contrast.

## 3.2 Performance metrics

In the proposed method, DSC and sensitivity were adopted as evaluation metrics to evaluate the performance of different methods in OCT image segmentation. These two metrics are used to assess the similarity between the segmentation results and the true annotations, as well as the pixel-level classification performance, respectively. The specific calculation formulas are as follows:

$$\text{DSC} = \frac{2\,|y_{\text{pred}} \cap y_{\text{true}}|}{|y_{\text{pred}}| + |y_{\text{true}}|} \tag{8}$$

$$\text{Sensitivity} = \frac{TP}{TP+FN} \tag{9}$$

In Equ. 8, $y_{\text{pred}}$ represents the predicted segmentation result, and $y_{\text{true}}$ represents the true value. In Equ. 9, $TP$ stands for true positive, and $FN$ stands for false negative.

## 3.3 Comparisons with state-of-the-art approaches

The EDU-Net model was compared with common models, such as FCN[17], U-Net [错误!未找到引用源。], Seg-Net[19], and Deeplabv3+[20], to preliminarily explore the performance of each model in terms of DSC and sensitivity and verify the effectiveness of EDU-Net. Subsequently, to further highlight the advantages of EDU-Net, several retinal anatomical structure segmentation models in recent years, as shown below:

•**MDAN-UNet:**[22] MDAN-UNet employs multiscale feature extraction and dual attention in a nested U-Net structure for effective local-global information fusion.

•**MsTGANet:**[24] MsTGANet utilizes the multiscale Transformer architecture to capture long-range dependencies and comprehensive context information.

•**CE-Net:**[31] CE-Net designs a context encoder network to extract multilevel semantic information while preserving spatial details.

- **CPF-Net:**[23] CPF-Net combines the context pyramid fusion module to effectively integrate multiscale context information.

- **Y-Net:**[32] Y-Net adapts a dual-branch network, with one branch extracting spectral features and the other capturing spatial information, to achieve complementary information fusion.

## 4. Results and analysis

### 4.1 Overall performance

#### 4.1.1. Results on in-house data

As shown in Table 1, in the IRF lesion segmentation task, the DSCs of EDU-Net were 5.11% and 3.04% higher than those of the best-performing Seg-Net and MsTGANet, respectively. However, in the SRF lesion segmentation task, the overall performance of EDU-Net was comparable to that of other models. Although its DSC index improved by 2.99% and 1.51% compared with those of the best-performing U-Net++ and Y-Net, respectively, it did not demonstrate a significant lead as it did in the case of IRF. In terms of sensitivity, EDU-Net generally had the second highest detection capabilities for both lesion types, only 3.1% lower than the best single-item performance Unet++ in the IRF index and a small gap lower than the best-performing Y-Net in the SRF index.

Fig. 8 shows the qualitative results of each model. IRF is usually presented as small and irregularly shaped scattered effusions, which increase the difficulty of segmentation. The first three rows show that EDU-Net improved the recognition of IRF regions, especially in the low-contrast lesion in the third row. CPF-Net and EDU-Net are better at completely capturing these targets than the models that are prone to missed detections. However, the SRF visualization segmentation results showed that EDU-Net had a certain degree of false-positive errors. These misjudged areas may be the areas where blood vessels pass through, which present similar features to SRF lesions under OCT imaging conditions. Therefore, these two methods for amplifying edge responses (CPF-Net and EDU-Net) may lead to false-positive errors.

### 4.1.2. Results on public data

Comparative experiments on the public RETOUCH dataset were conducted to further verify the effectiveness and robustness of EDU-Net.

Overall, EDU-Net achieved the best quantitative metrics under different manufacturers' OCT devices (with different imaging resolutions, noise levels, and contrasts), demonstrating its robustness across devices and datasets (Table 2). Specifically, in the quantitative DSC for IRF segmentation, EDU-Net achieved the scores of $79.54\pm0.52$, $76.55\pm1.86$, and $78.42\pm1.07$, outperforming the second-best models in each dataset by 2%, 3.95%, and 3.12%, respectively. These results comprehensively proved the significant advantages of EDU-Net in IRF lesion segmentation task and its stability and generalization ability under different clinical equipment conditions. However, similar to the results of the in-house dataset, the overall performance of EDU-Net was comparable to that of the other methods in terms of SRF segmentation performance. It achieved a relatively prominent DSC ($90.54\pm1.54$) in the Spectralis data volume only. For additional PED lesions (because its main manifestation is structural changes rather than a direct reflection of inflammatory or exudative activity), EDU-Net achieved the best results in DSC, which were $89.80\pm0.68$, $87.52\pm2.81$, and $86.01\pm1.39$, outperforming the second-best model by 1.08%, 3.82%, and 0.29%, respectively.

## 4.2. Ablation experiments

Ablation studies were conducted on EDU-Net variants to further explore the effectiveness of the key architectures. First, the necessity of introducing different key components was verified. Second, the EfficientNet branch was replaced with the U-Net structure to explore the balance between model performance and efficiency.

The experimental results in Tables 3 and 4 verified the following core assumption: fusing global and local branches can effectively address the challenges of complex and various morphological segmentation. Specifically, in the in-house dataset, when the global branch was used alone, the DSCs of IRF and SRF were 0.8003 and 0.8686, respectively; when the local branch was used alone, the DSCs were 0.7994 and 0.8360, respectively. After fusion, DSC increased to 0.8291 and 0.8884, an increase of approximately 3 or 4 percentage points. Similar improvements were observed on the Cirrus volume in RETOUCH, with the DSC for the three lesions increasing by 2.35%, 1.54%, and 2.51% after fusing the branches. This finding proved the effectiveness of the fusion strategy.

In the in-house dataset, the introduction of MC-EGA improved the IRF segmentation performance of the proposed model from 0.8291 to 0.8448, an increase of 1.57%, and the SRF segmentation performance improved by 1.96%. In the Cirrus volume in RETOUCH, the quantitative DSCs of the three types of lesions improved by 2.13%, 0.14%, and 0.56%. The results showed the contribution of the MC-EGA module to retinal lesion segmentation, especially for IRF.

Finally, after the EfficientNet branch (approximately 82.33 M) was replaced with U-Net (approximately 131.81 M), the metrics for both lesions in the in-house dataset decreased by approximately 3 or 4 percentage points. However, the verification results on the Cirrus volume data showed minimal difference. Overall, EfficientNet not only reduced the number of model parameters but also maintained high performance.

## 4.3 Parameter tuning

As shown in Fig. 9, when the weights returned to 1:1, the model achieved the best overall performance in terms of DSC and sensitivity indicators. However, as the weight disparity increased, the overall performance declined, further confirming that effective complementarity between branches can only be achieved when the weights are similar

## 4.4. Interpretive exploration of EDU-Net

First, the Grad-CAM activation maps[33] of shallow and deep fusion features of lesions were analyzed. As shown in Fig. 10, the Grad‑CAM activation maps of deep features reflected the network's attention to global semantic information, showing a broad and continuous response that can capture large-scale structures and patterns. By contrast, the activation maps of shallow features highlighted local details and high-frequency information, showing point-like or small fragment activations that can capture fine edges, textures, and local changes. Interpretability analysis indirectly reflected the necessity of fusing global and local features.

Second, the role of MC-EGA was explored. As shown in Fig. 11, the MC‑EGA module output exhibits lesion‑specific activation patterns. (a) The strongest activations occur in high‑density IRF regions. Given that no SRF lesions were present, these activation regions were continuously distributed along the retinal boundaries. (b) By fusing high‑frequency textures with deep semantic features, the module enhanced sensitivity to low‑contrast IRF, and an activation effect similar to a "box selection" was produced at the boundary of the SRF. (c) Although the sample did not contain SRF lesions, when the hot zones were continuously distributed along the retinal layers, the imaging features at the vascular courses were similar to those of SRF, which were continuously magnified in subsequent network layers, eventually leading to false positives.

# 5. Discussion

In this study, the EDU-Net DL framework was proposed, mainly for segmenting DME in retinal OCT images. Extensive experiments on in-house and public datasets demonstrated the robustness and accuracy of EDU-Net. In view of the complex and diverse morphology of retinal liquid lesions and the difficulty in effectively fusing local details and global context, a dual branch of global and local feature extraction was designed to extract and fuse multiscale information of liquid lesions. Tables 1 and 2 and the ablation experiments show the effectiveness of global–local feature dual-branch extraction and fusion. The MC-EGA module was introduced to optimize the segmentation of lesion edge areas to address the problem of fuzzy boundaries. Tables 1 and 2 and the ablation experiments show that MC-EGA performed well in the IRF lesion segmentation task. However, for SRF lesions, it did not achieve remarkable overall optimization. The visualization analysis in Figs. 8 and 11 shows that the MC-EGA module may misjudge some blood vessels as SRF lesions, thereby causing false-positive issues. Although the false-positive phenomenon in some images lowered the average index of SRF lesions, this module still made certain contributions to optimizing the edge features of SRF and IRF. Moreover, the false-positive problem can be alleviated or avoided by post-processing methods, so this module was retained.

This work has several limitations. First, although the MC-EGA module can increase the segmentation accuracy of IRF, it may cause false-positive errors for SRF. Second, fluid diffusion complies with the laws of physics, and the proposed method has not yet involved constraints in this regard. In the future, the authors intend to further explore this topic, including but not limited to (1) a universal and efficient fluid segmentation method; (2) a more effective high-frequency detail information extraction method and (3) an introduction of fluid physics constraints.

# Figures and Tables

Figure 1. Characteristics of retinal fluid in DME.

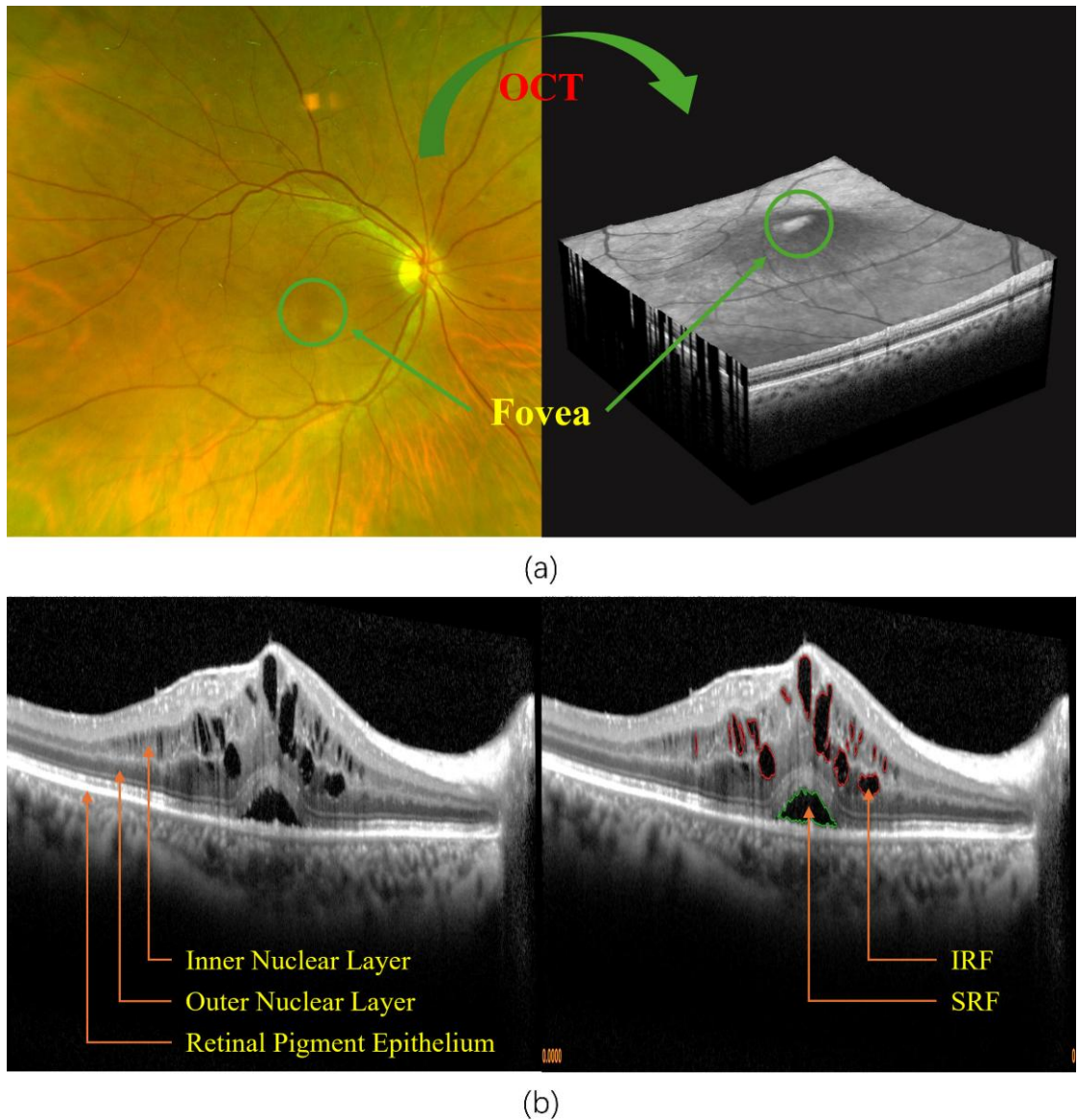

(a)

(b)

(a) Left: Posterior pole fundus photograph of a patient with DME; Right: OCT 3D imaging of the corresponding macular region. The green circle indicates the fovea. (b) B-scan image of the same DME patient. Left: The inner nuclear layer (INL), outer nuclear layer (ONL), and retinal pigment epithelium (RPE) are labeled; Right: Intraretinal fluid (IRF, red area) and subretinal fluid (SRF, green area) are marked.

Figure 2. Architecture diagram of the EDU-Net model.

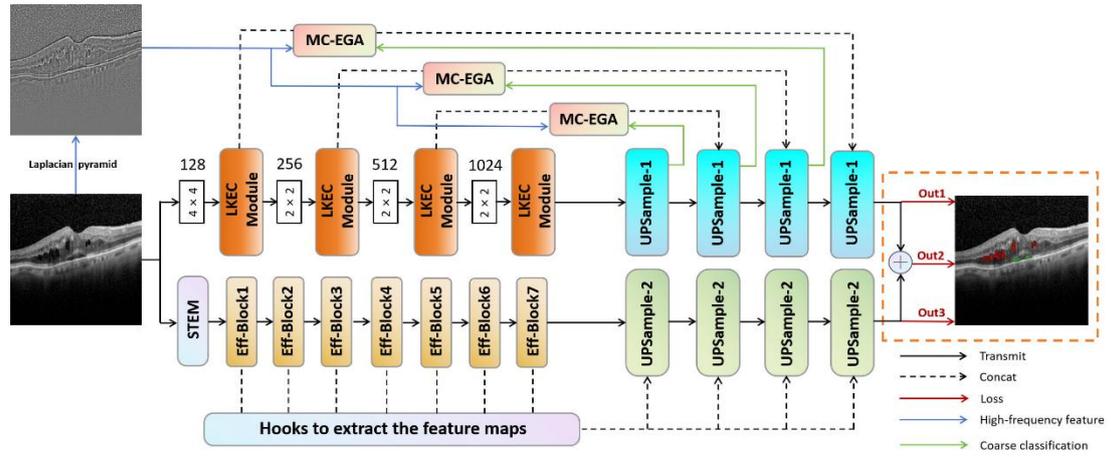

Figure 3. Framework diagram of the EfficientNet model.

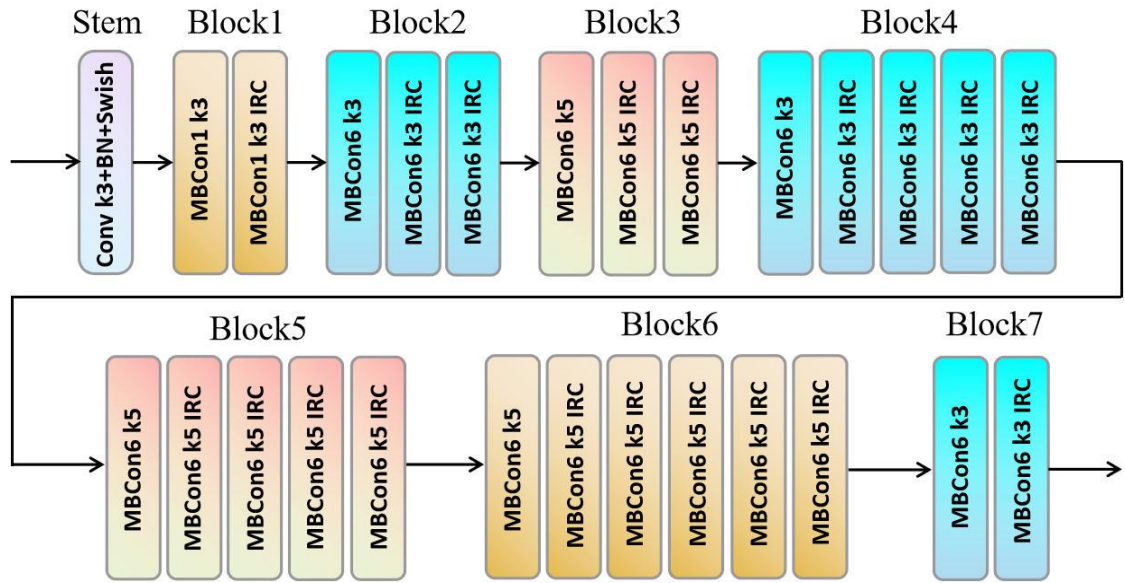

Figure 4. Architecture diagram of the MBConv module.

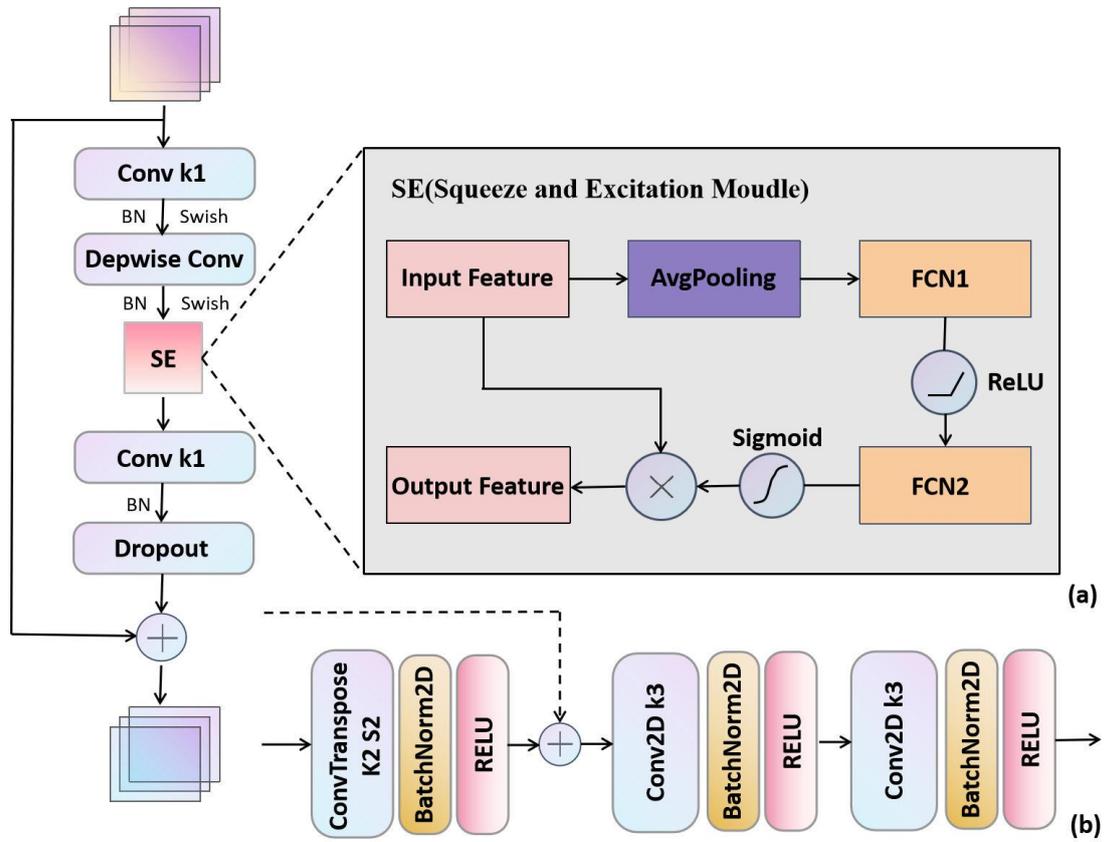

Figure 5. Architecture diagram of the LKEC module.

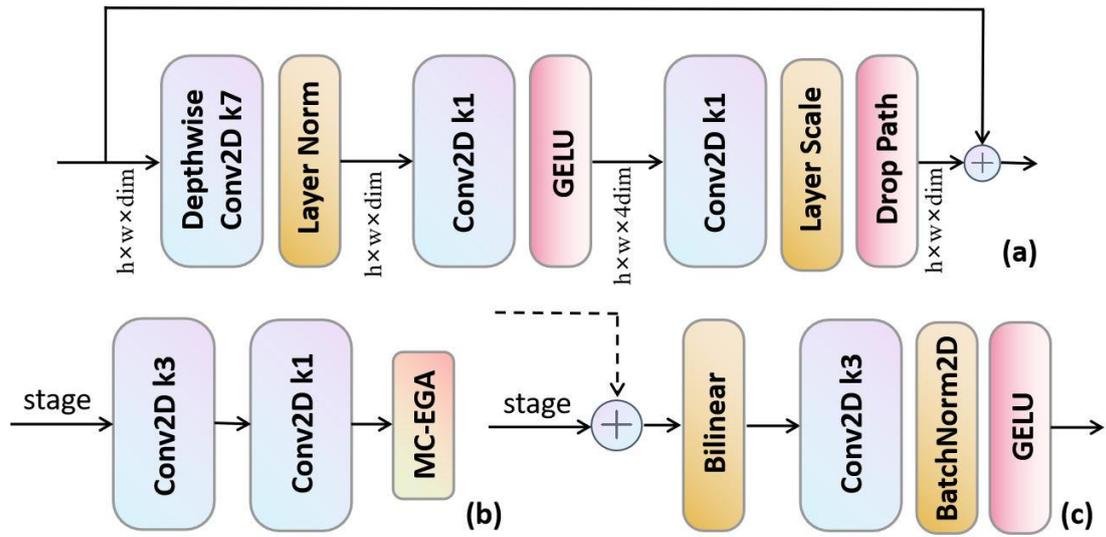

Figure 6. Architecture diagram of the MC-EGA module.

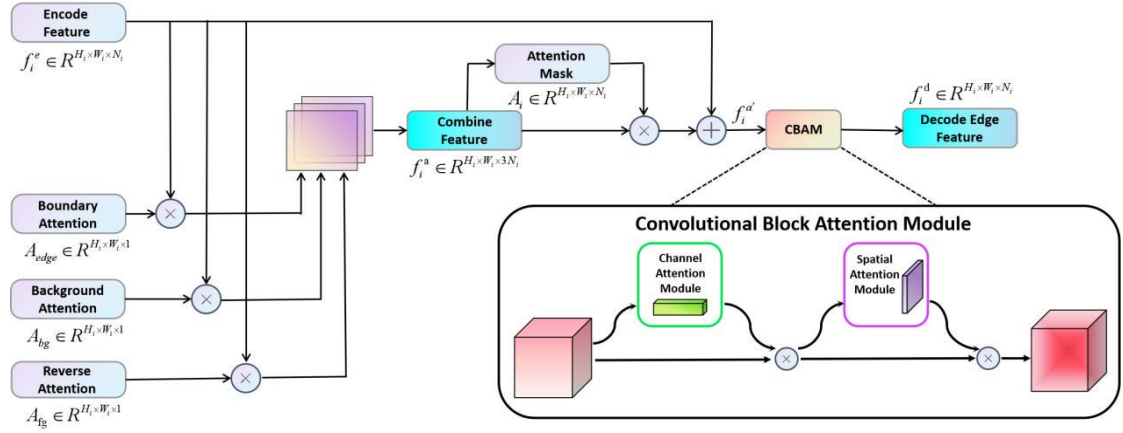

Figure 7. OCT images from three datasets.

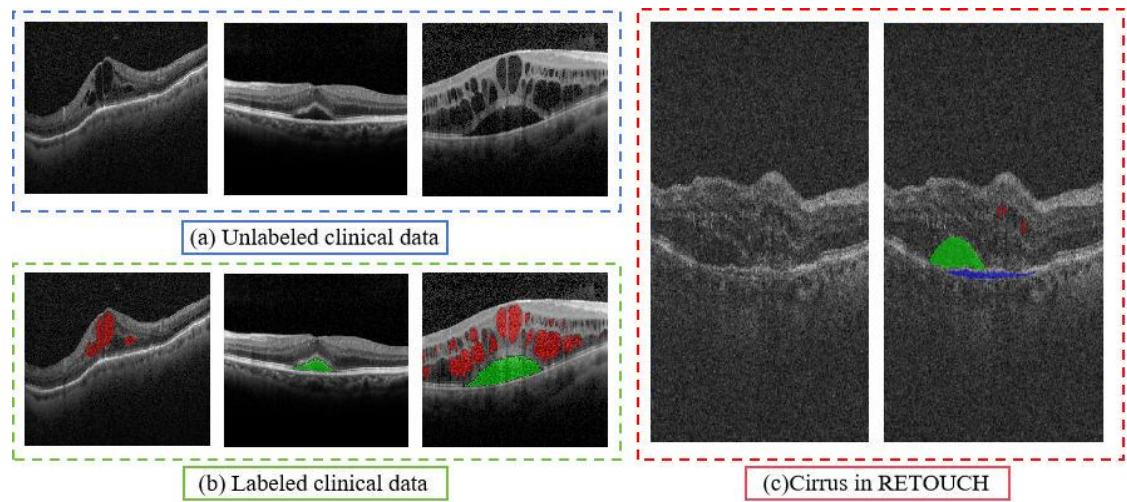

(a) Original DME OCT image from the Mendeley dataset; (b) OCT image annotated by ophthalmologists using Mimics software, with IRF marked in red and SRF marked in green; (c) Original and annotated OCT images from the RETOUCH Challenge dataset, with PED (pigment epithelial detachment) marked in blue.

Figure 8. Fluid segmentation results of different algorithms.

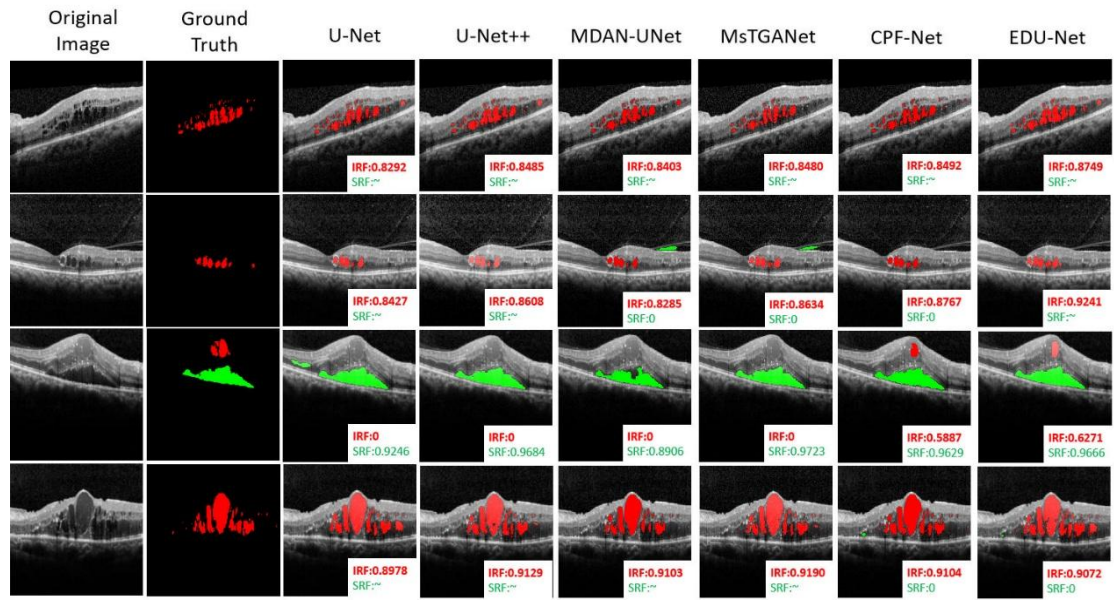

Figure 9. Hyperparameter tuning results.

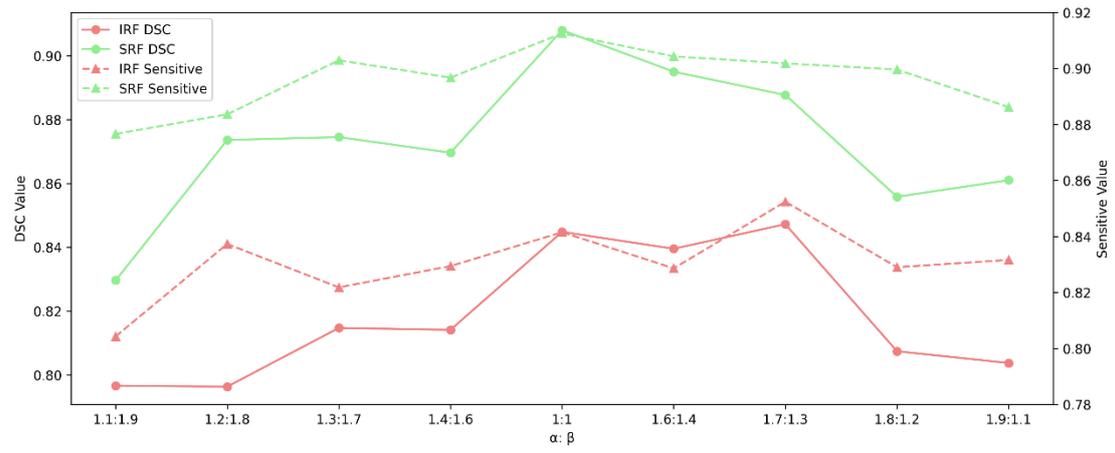

Figure 10. Grad-CAM heatmaps of two B-scan images. (a) and (b) Left: deep features; Right: shallow features.

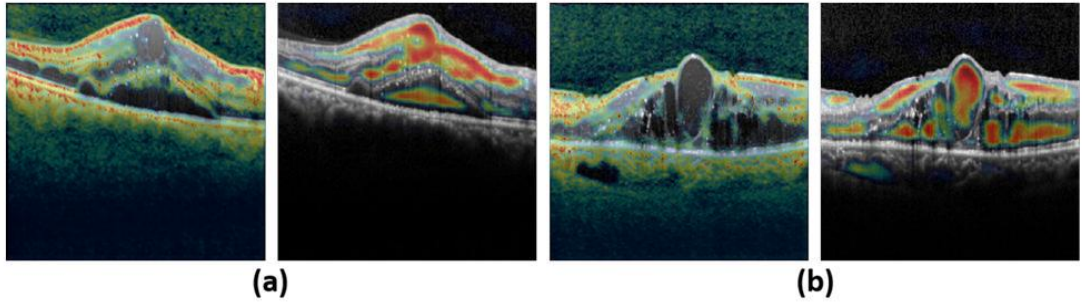

Figure 11. Output heatmaps of the MC-EGA module.

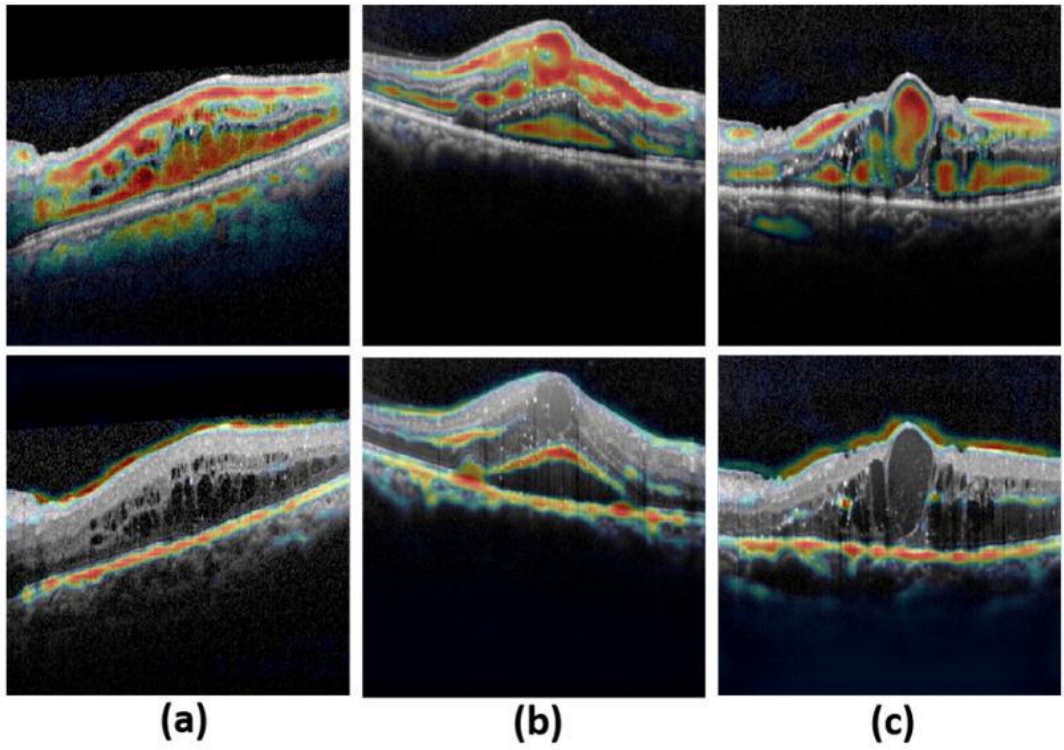

Table 1. Compares the performance differences of each algorithm in terms of DSC and sensitivity.

|  | DSC | | Sen | |
|---|---|---|---|---|
|  | IRF | SRF | IRF | SRF |
| FCN | 0.7461 | 0.7475 | 0.8226 | 0.8037 |
| U-Net | 0.7885 | 0.8688 | 0.7979 | 0.9012 |
| Seg-Net | 0.7937 | 0.8593 | 0.7929 | 0.8906 |
| Deeplabv3+ | 0.7375 | 0.7984 | 0.7489 | 0.8190 |
| Attention UNet[34] | 0.7838 | 0.8727 | 0.8019 | 0.9035 |
| TransUNet[35] | 0.7646 | 0.8626 | 0.6883 | 0.8062 |
| UNet++ | 0.7874 | 0.8781 | **0.8725** | 0.9084 |
| MDAN-UNet | 0.8001 | 0.8700 | 0.7766 | 0.8842 |
| MsTGANet | 0.8144 | 0.8859 | 0.8262 | 0.9088 |
| CE-Net | 0.7893 | 0.8680 | 0.7858 | 0.8939 |
| CPF-Net | 0.8039 | 0.8439 | 0.8207 | 0.9036 |
| Y-Net | 0.7944 | 0.8929 | 0.7899 | **0.9172** |
| EDU-Net | **0.8448** | **0.9080** | 0.8415 | 0.9125 |

Table 2. Five-fold cross-validation method based on DSC and sensitivity metrics in RETOUCH.

| Scanner | Metrics | Fluid | Network | | | | | | | |
|---|---|---|---|---|---|---|---|---|---|---|
| | | | U-Net | UNet++ | MDAN-UNet | MsTGANet | CE-Net | CPF-Net | Y-Net | EDU-Net |
| Cirrus | DSC | IRF | 75.08±0.98 | 75.59±1.64 | 71.66±1.33 | 75.00±1.61 | 77.54±1.21 | 76.78±0.77 | 75.08±2.59 | **79.54±0.52** |
| | | SRF | 89.14±1.48 | 89.33±1.10 | 86.26±2.45 | 88.63±0.43 | **90.92±0.38** | 90.29±0.59 | 88.77±1.25 | 90.68±1.31 |
| | | PED | 87.96±2.17 | 86.64±2.70 | 81.76±2.47 | 85.63±1.42 | 88.72±0.46 | 88.70±1.50 | 86.11±2.21 | **89.80±0.68** |
| | Sen | IRF | 76.50±1.10 | 76.99±2.20 | 72.44±2.87 | 76.95±2.09 | 78.21±2.29 | 76.31±1.78 | 75.47±2.72 | **80.19±0.42** |
| | | SRF | 91.03±1.92 | 90.23±2.45 | 85.73±3.27 | 90.58±1.38 | 91.99±0.80 | 90.92±1.03 | 89.06±2.23 | **92.14±0.77** |
| | | PED | 88.80±2.14 | 88.19±2.50 | 82.45±3.28 | 87.50±1.60 | 89.42±0.68 | 89.67±1.44 | 88.20±2.58 | **90.46±1.47** |
| Spectralis | DSC | IRF | 67.55±3.63 | 70.14±1.82 | 71.02±1.72 | 71.39±2.88 | 71.26±1.41 | 72.60±2.31 | 71.41±2.45 | **76.55±1.86** |
| | | SRF | 83.92±3.46 | 83.50±3.98 | 85.56±1.43 | 84.41±3.54 | 87.92±2.58 | 87.75±2.54 | 87.24±3.04 | **90.54±1.54** |
| | | PED | 79.16±3.15 | 79.64±1.52 | 80.16±2.75 | 79.49±1.98 | 83.55±2.28 | 83.70±0.49 | 80.04±2.64 | **87.52±2.81** |
| | Sen | IRF | 71.84±4.61 | 72.93±4.83 | 68.09±3.60 | 73.76±3.61 | 71.20±1.83 | 75.31±1.69 | 73.91±2.75 | **77.99±2.53** |
| | | SRF | 90.63±3.71 | 91.66±2.79 | 85.02±2.81 | 89.56±3.47 | 90.08±2.62 | **91.20±1.64** | 90.92±2.47 | 90.27±2.76 |
| | | PED | 81.19±3.16 | 80.84±4.11 | 83.00±3.53 | 86.41±1.89 | 87.04±3.27 | 86.25±2.45 | 85.37±5.83 | **89.01±2.34** |
| Topcon | DSC | IRF | 73.54±3.33 | 73.26±2.37 | 74.45±1.81 | 73.76±0.92 | 74.67±0.97 | 75.30±2.30 | 75.00±1.25 | **78.42±1.07** |
| | | SRF | 83.69±5.40 | 85.81±2.36 | 85.30±4.39 | 85.37±4.26 | 88.07±1.38 | 88.45±1.33 | 88.36±1.46 | **88.53±2.50** |
| | | PED | 81.98±1.02 | 83.50±2.73 | 77.44±3.28 | 80.95±3.48 | 85.72±1.77 | 85.62±1.25 | 84.04±1.64 | **86.01±1.39** |
| | Sen | IRF | 75.42±1.60 | 74.92±3.90 | 73.26±2.82 | 75.66±3.86 | 75.28±0.76 | 75.38±4.59 | 76.58±1.63 | **79.39±2.04** |
| | | SRF | 91.91±2.64 | 91.19±1.30 | 86.48±2.70 | 90.90±1.93 | 91.91±0.86 | 91.17±1.17 | **92.87±0.49** | 91.85±1.22 |
| | | PED | 85.17±1.41 | 85.00±3.39 | 76.87±3.89 | 82.57±2.87 | 86.93±1.92 | **87.68±1.11** | 85.19±1.77 | 86.80±1.59 |

The average and standard deviation were calculated (unit: percentile).

Table 3. Ablation results on clinical datasets.

| Dataset | Global branch | Local branch | MC-EGA | U-Net | DSC | |
| --- | --- | --- | --- | --- | --- | --- |
| | | | | | IRF | SRF |
| Clinical Dataset | ✔ | | | | 0.8003 | 0.8686 |
| | | ✔ | | | 0.7994 | 0.8360 |
| | ✔ | ✔ | | | 0.8291 | 0.8884 |
| | ✔ | | ✔ | ✔ | 0.8153 | 0.8596 |
| | ✔ | ✔ | ✔ | | **0.8448** | **0.9080** |

Table 4. Ablation results on Public datasets.

| Dataset | Global branch | Local branch | MC-EGA | U-Net | DSC | | |
|---|---|---|---|---|---|---|---|
| | | | | | IRF | SRF | PED |
| RETOUCH (Cirrus) | ✔ | | | | 75.06±1.15 | 89.00±1.66 | 86.73±2.42 |
| | | ✔ | | | 74.69±0.74 | 88.73±1.44 | 86.71±1.99 |
| | ✔ | ✔ | | | 77.41±1.13 | 90.54±0.96 | 89.24±1.46 |
| | ✔ | | ✔ | ✔ | 78.82±1.49 | **90.88±0.85** | 89.13±2.07 |
| | ✔ | ✔ | ✔ | | **79.54±0.52** | 90.68±1.31 | **89.80±0.68** |